\documentclass[11pt]{article}
\usepackage{moriond,epsfig}

\def\be{\begin{equation}}
\def\ee{\end{equation}}
\def\bea{\begin{eqnarray}}
\def\eea{\end{eqnarray}}

\begin{document}
\title{DECOHERENCE FROM VACUUM FLUCTUATIONS}

\author{\underline{MARKUS B\"UTTIKER}}

\address{D\'epartement de Physique Th\'eorique,
Universit\'e de Gen\`eve, CH-1211 Gen\`eve 4, Switzerland}

\maketitle\abstracts{Vacuum fluctuations are a source of 
irreversibility and decoherence. 
We investigate the persistent current and its fluctuations in 
a ring with an in-line quantum dot with an Aharonov-Bohm 
flux through the hole of the ring. The Coulomb blockade leads 
to persistent current peaks at values of the gate 
voltage at which two charge states of the dot 
have the same free energy. We couple the structure to an external 
circuit and investigate the effect of the zero-temperature 
(vacuum fluctuations) on the ground state of the ring. 
We find that the ground state of the ring undergoes a crossover 
from a state with an average persistent current much larger than 
the (time-dependent) mean squared fluctuations 
to a state with a small average persistent current 
and large mean squared fluctuations. We discuss the spectral 
density of charge fluctuations and 
discuss diffusion rates for angle variables characterizing 
the ground state in Bloch representation. 
}

\section{Introduction}

In this work we are interested in the coherence properties 
of the ground state of a mesoscopic system coupled to an environment. 
In the zero-temperature limit, the only source of decoherence 
are then provided by vacuum fluctuations.  
The work is motivated by a recent discussion 
in the mesoscopic physics community which largely 
insists that dephasing rates tend to zero (typically with some 
power law) as a function of temperature and that there is therefore
no dephasing in the zero-temperature limit. For references to this 
discussion we direct the reader to Ref. \cite{natel} .  

A key argument is that in the zero-temperature limit a system 
can not excite a bath by giving away an energy quantum nor can 
a bath in the zero-temperature limit give an energy quantum 
to the system. This view holds that dephasing is necessarily 
associated with an energy transfer (a real transition) 
and since this is impossible there is in the zero-temperature 
limit no dephasing. However, 
this argument rests on the assumption that the system and the bath 
are in their ground state which they assume in the absence of 
any coupling. In the ground state of the system + bath, 
even at absolute zero, the energy of the small system 
fluctuates \cite{cedr,naga} and the ground state of the small 
system is not a pure state. 

\begin{figure}
\vspace*{-0.5cm}
\epsfysize=7cm
\epsfxsize=5cm
\centerline{\epsffile{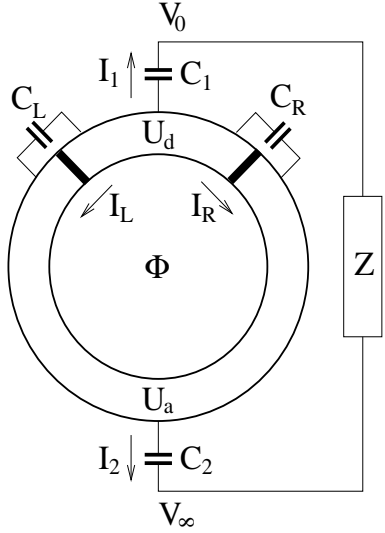}}
\vspace*{0.3cm}
\caption{\label{system} Ring with an in-line dot subject to a 
flux $\Phi$ and capacitively coupled to an external impedance $Z$.
}
\end{figure}

To illustrate this we investigate a simple mesoscopic system, 
a quantum dot with its leads formed into a ring \cite{cedr},  
as shown in Fig. \ref{system}. 
Such a ring, when the coupling between the dot and the arms 
of the ring is sufficiently weak, exhibits 
Coloumb blockade peaks in the persistent current
at gate voltages which equalize the free energy 
of the $N$-th and $N+1$-th charge state of the dot. 
We couple this system capacitively to an external circuit 
with ohmic resistance $R$ and investigate
the persistent current near a resonance. 
At absolute zero temperature the gate voltage fluctuates 
due to vacuum fluctuations of the resistor.  
We show that the ground state of this system is not a pure coherent
state. 
In the two state limit, the system shown in Fig. \ref{system} 
maps onto the spin boson problem: for this system there exits a large 
literature \cite{legg,weiss}. 
We emphasize that the work presented here is not directly related to
weak localization but considers the persistent current (a ground state
property). 

\section{Coulomb Blockade Peaks of the Persistent Current}

Consider the system shown in Fig. \ref{system}. 
The arm of the ring contains electrons in 
levels with energy $E_{am}$ and the dot contains electrons 
in levels with energy $E_{dn}$. First let us for a moment
neglect tunneling. Let $F_{N}$ be the free energy 
for the case that there are $N$ electrons in the dot and $M$
electrons in the arm. The transfer of an electron from the 
arm to the dot gives a free energy $F_{N+1}$.
The difference of these two free energies 
is
\begin{equation}
\hbar \epsilon_{0}= F_{N+1}-F_{N} = E_{d(N+1)}-E_{aM} + 
\frac{e^{2} (N + 1/2 - C_{0}V_{e})}{C} . 
\end{equation}
Here the first two terms arise from the difference in kinetic energies.
The third term results from the charging energy of the dot.
$C^{-1}_{0} = C^{-1}_{i} + C^{-1}_{e}$ is the series capacitance
of the internal capacitance $C_{i} = C_{L} + C_{R}$ 
and the external capacitance $C^{-1}_{e} = C^{-1}_{1} + C^{-1}_{2}$.
The total capacitance is  $C = C_{i} + C_{e}$. 
Tunneling through the barriers connecting the dot and 
the arm is described by amplitudes $t_{L}$ and $t_{R}$
and depends on the Aharonov-Bohm flux $\Phi$ in 
the following way \cite{mbcs}, 
\begin{equation}
\label{Delta0}
{\hbar \Delta_0 \over 2}
= \left(t_L^2 + t_R^2 \pm 2 t_L t_R \cos{2\pi \Phi \over \Phi_0}\right)^{1/2},
\end{equation}
where $\Phi_0 = hc/e$ is the single electron flux quantum. 
The sign depends on the number of electrons in the 
dot and arm: it is positive if the total number is odd, and it is 
negative if the total number is even. 
The voltage across the system is $V = V_{0} - V_{\infty}$. 
The Hamiltonian consists of the system part $H_{0}$,
the coupling $H_{c}$ and the bath $H_{B}$ with 
\begin{equation}
H_{0} = \frac{\hbar \epsilon_{0}}{2} \, \sigma_{z} - 
\frac{\hbar \Delta_{0}}{2} \, \sigma_{x} , 
\,\,\,\, H_{c} = \frac{C_{0}}{C_{i}} \frac{eV}{2} \, \sigma_{z} .
\label{twostate1}
\end{equation}
$H_{0}$ has eigenstates with energies
\begin{equation}
E_{\mp} = \mp
\frac{\hbar}{2}  \sqrt{\epsilon_{0}^{2} + \Delta_{0}^{2}}
\equiv  \mp \frac{\hbar \Omega_{0}}{2} . 
\end{equation}
In the presence of a flux $\Phi$ the ground state of 
the ring-dot system (see Fig. \ref{system})
permits a persistent current which is \cite{mbcs}  
\begin{equation}
<I> = -c \frac{dE_{-}}{d\Phi} = \mp \, {e}
\frac{4 \pi t_{L} t_{R}} {\Omega_{0}} \, 
\sin(2\pi \Phi/\Phi_{0}). 
\end{equation}
The equilibrium current is a pure quantum effect: only 
electrons whose wave functions are sufficiently 
coherent to reach around the loop contribute 
to the persistent current. Thus the persistent current is
a measure of the coherence of the ground state. 
At resonance $\epsilon = 0$ the current is of the 
order of $et$ with $t$ a transmission amplitude and 
it decreases and becomes of the order 
of $et^{2}/\epsilon $ as we move away from resonance.  

\begin{figure}
\vspace*{-4cm}
\epsfysize=9cm
\epsfxsize=7cm
\centerline{\epsffile{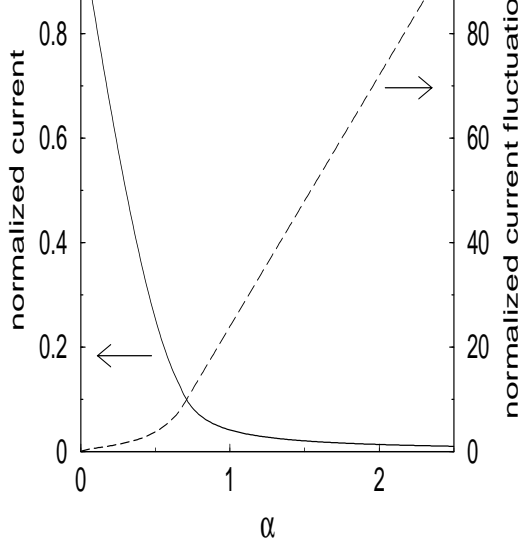}}
\caption{\label{kondopc} The persistent current $<I>$ at resonance
$\epsilon = 0$ for a symmetric ring (solid line) 
and the fluctuations of the
circulating current $(<(I (t) -  <I (t)>)^{2}>)^{1/2}/<I>$
(dashed line)
as a function of the coupling parameter $\alpha$.   
The average persistent current is in units of the persistent 
current at $\alpha = 0$. The fluctuations are normalized by the average 
current. The parameters are $\omega_c = 25 \Delta_0$.
After \protect\cite{cedr}.}
\end{figure}

In the two-level limit of interest here the transmission 
amplitudes $t_L$ and $t_R$ are taken to be very small 
compared to the level spacing in the dot and in the arm. 
For larger transmission amplitudes the system without 
a bath will already exhibit a Kondo effect \cite{eckle,kang,affl}. 
A completely equivalent model consists of a 
superconducting electron box \cite{bouch} 
which can be opened to admit an Aharonov-Bohm flux \cite{moji,makh}.

We are interested in the effect of the bath on the persistent 
current. With the bath included our two-level system 
becomes a spin-boson problem \cite{legg,weiss}.  
Cedraschi, Ponomarenko and the author \cite{cedr,annals} used 
known solutions based on a Bethe ansatz 
and perturbation theory (for the anisotropic Kondo model) 
to provide an answer. 
For the symmetric case $t_{L} = t_{R}$, 
$C_{L} = C_{R}$,  the average current 
is shown in Fig. \ref{kondopc} as a function 
of the coupling parameter 
\begin{equation}
\label{Gamma}
\alpha \equiv {R \over R_K} \, \left( {C_0 \over C_i} \right)^2 , 
\,\,\,\,\,  \Gamma \equiv \pi \alpha {\Delta_0^2 \over \Omega_0}
\end{equation}
with $R_K \equiv h/e^2$ the quantum of resistance. 
We have also introduced the relaxation rate $\Gamma$ which as will
be shown governs the zero-temperature decoherence of the system. 
The persistent current 
is in units of the current for $\alpha = 0$. 
In addition Ref. \cite{cedr} also 
investigated the (instantaneous) fluctuations 
of the equilibrium current away from its average, 
$(<(I (t) -  <I (t)>)^{2}>)^{1/2}$ shown in Fig. \ref{kondopc} 
in units of the average current $<I(t, \alpha )>$.  
With increasing resistance we have thus a {\it cross over} from a state 
with a well defined persistent current (small mean-squared fluctuations)
to a fluctuation dominated state in which the mean-squared fluctuations 
of the persistent current are much larger than the average 
persistent current. For the derivation we refer the reader to 
Refs. \cite{cedr} and \cite{annals}. Here we pursue a discussion 
based on Langevin equations \cite{lange}.  
This approach is valid only for weak coupling constants $\alpha \ll 1$
but has the benefit of being simple. 

We want to find the time evolution of a state $\psi(t)$
of the two-level system in the presence of the bath. 
We write the state of the two-level system 
\begin{equation}
\label{state}
\psi = e^{i\chi/2}
{\cos{\theta \over 2} \, e^{i\varphi/2} \hfill
\choose
\sin{\theta \over 2} \, e^{-i\varphi/2}},
\end{equation}
with $\theta$, $\varphi$ and $\chi$ real.  This is the most general
form of a normalized complex vector in two dimensions.  In terms of
$\theta$, $\varphi$ and the global phase $\chi$, the time dependent
Schr\"odinger equation reads
\begin{eqnarray}
\label{central1}
\dot{\varphi} &=& -\varepsilon_{0} - \delta \varepsilon (t) 
- \Delta_0 \cot\theta \cos\varphi, \\
\label{central2}
\dot{\theta} &=& -\Delta_0 \sin\varphi, \\
\label{chideq}
\dot{\chi} &=&  \Delta_0 {\cos\varphi \over \sin\theta}.
\end{eqnarray}
Here $\delta \varepsilon (t) =  (e/h) (C_{0}/C_{i}) V (t)$
arises from the gate voltage fluctuations. 
As shown by Eq.~(\ref{chideq})
the dynamics of the phase $\chi$ is
completely determined by the dynamics of the phases $\theta$ 
and $\varphi$ and has no back-effect 
on the evolution of $\theta$ and  $\varphi$.
While $\chi$ is irrelevant for expectation
values, like the persistent current or the charge on the dot,  it
plays an important role, in the discussion of phase diffusion times.

To close the system of equations,  
we have to find the voltage which drops across the system. 
The charge transfer between the dot and its arms 
permits a displacement current 
through the system which we have to include to find the voltage
fluctuations. 
The charge operator on the dot for our problem is 
$\hat{Q}_d = (1/2) (\sigma_z + 1)$. 
Its quantum mechanical expectation value is 
$
{Q}_d = \langle \psi(t) | \hat{Q}_d
| \psi(t) \rangle = e \, \cos^{2}(\theta /2) .
$
The displacement current is proportional to the time-derivative of 
this charge, 
$ \dot{Q}_d = -{e \over 2} \sin\theta \, \dot{\theta}$
multiplied by a ratio of capacitances which has to 
be found from circuit analysis \cite{lange}. 
We find that the total current through the system is now 
given by $C_0  \dot{V} -
({C_0 /C_i}) ({e /2}) \sin\theta \, \dot{\theta}$. 
Using this result we find from conservation  
all currents that the fluctuating voltage 
across the system is determined by 
\begin{equation}
\label{central3}
V = - C_0 R \,  \dot{V} -
R \, {C_0 \over C_i} \, {e \over 2} \, \sin\theta \, \dot{\theta}
+ R I_{N}(t) .
\end{equation}
Eqs.~(\ref{central1},~\ref{central2}) and
(\ref{central3}) form a closed system of equations in which the
external circuit is incorporated in terms of an ohmic resistor $R$ 
in parallel with a fluctuating current
$I_{N}(t)$ with spectral density $S_{II}(\omega) = 
(\hbar \omega/R) coth(\hbar \omega/2kT)$. 
In the next section, we
investigate Eqs.~(\ref{central1},~\ref{central2}) and
(\ref{central3}) to find the effect of zero-point fluctuations on
the persistent current of the ring.

\section{Fluctuations around the ground state}
\label{expansion}

In the absence of the
noise term $I_{N}(t)$, the stationary states of the system of
differential equations, Eqs.~(\ref{central1},
\ref{central2}) and (\ref{central3}) 
are given by $\varphi \equiv \varphi_0$, with
$\varphi_0 = 0$ or $\varphi_0 = \pi$ and $\theta \equiv \theta_0$,
with $\cot\theta_0 = \pm {\varepsilon_0 \over \Delta_0}$. 
The lower sign applies for $\varphi_0 = 0$.  This is the {\it ground state}
for the ring-dot system at fixed $\varepsilon(t) \equiv
\varepsilon_0$, and the upper sign holds for $\varphi_0 = \pi$.
The energy of the ground state is $-\hbar\Omega_0/2$, thus the global
phase is $\chi_0(t) = \Omega_0 t$. 
We also introduce the ``classical'' relaxation time 
$\tau_{RC} \equiv RC_0$.

Now, we switch on the noise $I_{N}(t)$.  We seek $\varphi(t) ,
\theta(t)$, $\chi(t)$ and $V(t)$ in linear order in the noise
current $I_{N}(t)$.  We expand $\varphi(t)$ and $\theta(t)$ to
first order around the ground state, $\varphi = 0$ and $\theta =
\theta_0$. For $ \delta \varphi(t) = \varphi(t) - \varphi_0$,
$\delta\theta(t) = \theta (t) - \theta_0$, etc., 
we find in Fourier space,
\begin{eqnarray}
\label{central1linF}
-i \omega \delta\varphi &=& -\delta\varepsilon
+ {\Omega_0^2 \over \Delta_0} \, \delta\theta, \\
\label{central2linF}
-i \omega \delta\theta &=& -\Delta_0 \, \delta\varphi, \\
\label{central3linF}
-i \omega \, \delta \varepsilon
&=& {1 \over \tau_{RC}} \left[
-\delta\varepsilon - \Gamma \, \delta\varphi
+ {e \over \hbar} \, R \, {C_0 \over C_i} \, I_{N}
\right],\\ 
-i\omega \delta\chi &=& \Omega_0 \, {\varepsilon_0 \over \Delta_0}
\delta\theta.
\end{eqnarray}
Here we have also expanded 
the global phase $\chi(t)$ around its
evolution in the ground state $\chi_0(t) = \Omega_0 t$, and define
$\delta\chi(t) = \chi(t) - \chi_0(t)$.  
We note that there is no effect of the global shift in energy,
$\hbar\nu(t)$, as it is quadratic in the voltage $\delta V$, and we
are only interested in effects up to linear order in $\delta V$.

\section{Mapping onto a harmonic oscillator} 

Let us assume that the charge relaxation time of 
the external circuit $\tau_{RC}$ is very short compared to the dynamics 
of the two-level system $\tau_{RC} \ll \Omega_{0}$.
Eliminating $\delta\theta$ with the help 
of Eq. (\ref{central2linF}) and  $\delta\varepsilon$
with the help of Eq. (\ref{central3linF}) we find 
\begin{eqnarray}
\label{oscillator}
(\omega^{2} - i \omega  \Gamma - \Omega_0^2 ) \, \delta\theta
= \Delta_0 \, {e \over \hbar} \, {C_0 \over C_i} \, R \, I_{N} .
\end{eqnarray}
Thus we have mapped the dynamics of the fluctuations 
away from the ground state of this two-level system 
on the quantum Langevin equation of a damped harmonic oscillator. 
$\delta\theta$ plays the role of the charge,  
$\delta\varphi$ the role of the current 
and $\Gamma$ (defined in Eq. (\ref{Gamma}))
takes the role of the friction constant.
The spectral density $S_{\theta\theta}(\omega)$
is just that of the coordinate of the harmonic oscillator, 
\begin{equation}
S_{\theta\theta} (\omega)
= {2\pi \, \alpha \, \Delta_0^2 \, | \omega |
\over
\left[
\left( \omega^2 - \Omega_0^2 \right)^2
+ \Gamma^2 \omega^2 
\right]}.
\end{equation}
In the literature it is often 
the correlation function of $\sigma_z$ which is of interest. 
We have $<\psi | \sigma_z |\psi> = \cos(\theta)$
and thus for the fluctuations away from the average 
$\Delta <\psi | \sigma_z |\psi> = $$ - \sin(\theta_{0}) \delta \theta$.
Since $\sin(\theta_{0}) = \Delta_{0}/\Omega_{0}$, 
we find in the zero-temperature limit
$S_{\sigma_{z}\sigma_{z}} (\omega) = $
$(\Delta^{2}_{0}/\Omega^{2}_{0})
S_{\theta\theta} (\omega)$. 
This result agrees with an expression given by 
Weiss and Wollensak \cite{ww} and G\"orlich et al. \cite{gsw}
who have used an entirely different approach. 
For non-zero temperatures Weiss and Wollensak find in addition
a Debye peak around zero-frequency. 
The essential 
point is that the 
peaks are broadened with a relaxation rate $\Gamma$. 
Using Eqs. (\ref{central2linF}) and (\ref{central3linF})
we obtain similarly the spectral densities $S_{\varphi \varphi}(\omega)$, 
$S_{\chi\chi}(\omega)$ and cross-correlations like 
$S_{\theta \varphi}(\omega)$.

\section{Suppression of the Persistent Current}
\label{pc}
Let us next examine the reduction of the persistent current 
using the approach outlined above. 
We consider only the case 
of a symmetric ring $t_{R} = t_{L} \equiv t$ 
and $C_R = C_ L$ (see Fig. \ref{system}). 
The persistent current is the quantum and statistical 
average of the operator $\hat{I}_c = {\cal J} \sigma_{x}$
where $\cal J$ is given by
$
{\cal J}
= {\hbar c} {\partial \Delta_0 \over \partial \Phi}.
$
In general, in the non-symmetric
case, the operator for the persistent
current depends also on the capacitances
(see Appendix B of Ref. \cite{lange}). 
The quantum mechanical 
expectation value of the persistent current for the state given in
Eq.~(\ref{state}) reads
$
I(t) \equiv \langle \psi(t) | \hat{I}_c | \psi(t) \rangle
= {1 \over 2} {\rm Re}\, ( {\cal J} \sin\theta \, e^{-i\varphi}).
$
We are interested in the {\em statistically averaged} 
persistent current $\langle I(t) \rangle$.  
Therefore, we have to calculate the correlator $\langle \sin\theta \,
e^{-i\varphi} \rangle$.  Taking into account that for 
a harmonic oscillator the fluctuations are Gaussian, we find 
$ \langle \sin\theta e^{-i\varphi} \rangle
= \langle \sin\theta \rangle
\langle e^{-i\varphi} \rangle $. 
and 
$
\langle e^{-i\varphi(t)} \rangle
= e^{-i\varphi_0} \langle e^{-i\delta\varphi(t)} \rangle
= \langle
\exp ( -{\delta\varphi^2(t) / 2})
\rangle ,
$
where we have used that $\varphi_0 = 0$. In the weak coupling limit,
and in the extreme quantum limit,
$T=0$, we find for the time averaged mean-squared fluctuations
to leading order in $\Gamma$, 
\begin{equation}
\label{msdphi2}
\langle \delta\varphi^2(t) \rangle
= \int_0^{\omega_c} \frac{d\omega}{\pi} 
S_{\varphi\varphi}(\omega)
= \frac{\Omega_0}{\Delta_0^{2}} 
[ 2 \Gamma \ln {\omega_c \over \Omega_0} - \Gamma + \pi \Omega_{0} ]
\approx 2\alpha
\ln {\omega_c \over \Omega_0} . 
\end{equation}
Here we have assumed that the cut-off frequency is so large 
that the logarithmic term dominates. 
In the limit $\omega_c \gg \Omega_0$, we
can neglect $\langle \delta\theta^2(t) \rangle $ 
against $\langle
\delta\varphi^2(t) \rangle$.  We insert $\langle \delta\varphi^2(t)
\rangle$ and $\sin\theta_0 = \Delta_0/\Omega_0$ into $\langle
\sin\theta \, e^{-i\varphi} \rangle$, and find 
a noise averaged persistent current in the ring given by 
\begin{equation}
\label{pcsimple}
\langle I(t) \rangle
= -{\hbar c \over 2} {\partial \Delta_0 \over \partial \Phi}
{\Delta_0 \over \Omega_0}
\left( {\Omega_0 \over \omega_c} \right)^\alpha.
\end{equation}
The weak coupling limit corresponds to 
$\alpha \ll 1$. 
The power law for the persistent current obtained in
Eq.~(\ref{pcsimple}), as well as the exponent $\alpha$,
Eq.~(\ref{Gamma}) coincide in this limit 
with the result obtained by Cedraschi et al. \cite{cedr}  using a Bethe
ansatz solution. We next characterize the fluctuations of the
ring-dot subsystem in more detail.

\section{Phase Diffusion Times}
\label{dephasing}

Due to the vacuum fluctuations the ground state of the 
two-level system is a dynamic state. 
To see this we project the actual state of the 
system on the 
{\em ground state\/} $\psi_- = (\cos\theta_0/2, \sin\theta_0/2)$, with
eigenvalue $-\hbar\Omega_0/2$, and the {\em excited state\/} $\psi_+ =
(-\sin\theta_0/2, \cos\theta_0/2)$ with eigenvalue $\hbar\Omega_0/2$
of the isolated system. 
Instead of the wave function $\psi(t)$ it is more convenient 
to consider $\psi_R(t) \equiv
\exp(i\hat{H}_0 t/\hbar) \psi(t)$. 
To first order in $\delta\varepsilon$, we find for the wave function 
$\psi_R(t) = ( 1 + i c_{-}) \psi_{-} + c_{+} \psi_{+} e^{i\Omega_0 t} $
with 
\begin{equation}
\label{cplus}
c_{-} = {\delta\chi(t) \over 2}
- {\varepsilon_0 \over \Omega_0} {\delta\varphi(t) \over 2} , \,\,\,\, 
c_{+} = \left[ {\delta\theta(t) \over 2}
- i{\Delta_0 \over \Omega_0} {\delta\varphi(t) \over 2} 
\right] e^{i\Omega_0 t} .
\end{equation}
Expressing the fluctuations of the angle variables in terms 
of their fluctuation spectra, we find 
$\left\langle \left| c_{\mp} (t) - c_{\mp}(0) \right|^2 \right\rangle =
t/\tau_{\mp}$ with diffusion times \cite{lange} 
\begin{equation}
\label{tauplus}
\tau_- = {\hbar \over 2\pi\alpha \, kT}
{\Omega_0^2 \over \varepsilon_0^2},  \,\,\,\, 
\tau_+ = {1 \over \Gamma}
\tanh {\hbar\Omega_0 \over 2kT}.
\end{equation}
Note that $\tau_-$ 
depends on the detuning 
$\varepsilon_0$. In particular,
at resonance $\varepsilon_0=0$, the phase diffusion time $\tau_-$
diverges for any temperature.
The long time behavior of $\tau_{+}$,
is determined by frequencies near $\Omega_0$.  
Eq.~(\ref{tauplus}) holds for finite temperatures as
well as in the quantum limit. In the low-temperature or
quantum limit, however, $\tau_{+}$ {\em saturates } to a value
$1/\Gamma$. Thus at short and intermediate 
times the ground state is not a coherent state 
but exhibits diffusion. The dephasing rate \cite{ww,gsw,grif}
is one-half of the sum of the two rates $1/\tau_{\mp}$.


\section*{References}
\begin{thebibliography}{99}

\bibitem{natel}    D. Natelson, et al., 
                   Phys. Rev. Lett. {\bf 78}, 1821 (2001).    

\bibitem{cedr}     P. Cedraschi, V.~V. Ponomarenko, and M. B\"uttiker, 
                   Phys. Rev. Lett. {\bf 84}, 346  (2000).


\bibitem{naga}     K. Nagaev and M. B\"uttiker, (unpublished). 


                   
\bibitem{legg}     A. J. Leggett, et al.,  
                   Rev. Mod. Phys. {\bf 59}, 1 (1987).                    

\bibitem{weiss}    U. Weiss, 
                   {\it Quantum Dissipative Systems}, (Word Scientific, 2000).


\bibitem{mbcs}     M. B\"uttiker and C.~A. Stafford, 
                   Phys. Rev. Lett. {\bf 76},  495  (1996).

\bibitem{eckle}    H.-P. Eckle, H. Johannesson, and C.~A. Stafford, 
                   J. Low Temp. Phys. {\bf 118}, 475 (2000).  
                   cond-mat/0010101   

\bibitem{kang}     K. Kang and S.-C. Shin, Phys. Rev. Lett. {\bf 85}, 
                   5619 (2000).                   
                                    
\bibitem{affl}     I. Affleck and P. Simon, Phys. Rev. Lett. 
                   {\bf 86}, 2854 (2001); 
                   P. Simon, I. Affleck, cond-mat/0103175 .  

\bibitem{bouch}    V. Bouchiat, et al., 
                   J. of Superconductivity, {\bf 12}, 789 (1999). 
                   
                   
\bibitem{moji}     J. E. Moji et al, Science {\bf 285}, 1036 (1999). 
                   L. Tian, et al., 
                   in "Quantum Mesoscopic Phenomena and Mesoscopic Devices
                   in Microelectronics", 
                   edited by I. O. Kulik and R. Ellialtioglu, 
                   (Kluwer, Netherlands, 2000). p. 429. 


\bibitem{makh}     Y. Makhlin, G. Sch\"on, and A. Shnirman, 
                   in  "Quantum Physics at Mesoscopic Scale"
                   edited by D.C. Glattli, M. Sanquer and 
                   J. Tran Thanh Van
                   (EDP Sciences, Les Ulis, 2000). p. 113              


\bibitem{annals}   P. Cedraschi and M. B\"uttiker, 
                   Annals of Physics, {\bf 289}, 1 - 23 (2001). 
                  

\bibitem{lange}    P. Cedraschi and M. B\"uttiker, 
                   Phys. Rev. B {\bf 63}, 165312 (2001).

\bibitem{ww}       U. Weiss and M. Wollensak, 
                   Phys. Rev. Lett. {\bf 62}, 1663 (1989). 

\bibitem{gsw}      R. G\"orlich, M. Sassetti, and U. Weiss, 
                   Europhys. Lett. {\bf 10}, 507 (1989). 

\bibitem{grif}     M. Grifoni, E. Paladino and U. Weiss, 
                   Eur. Phys. J. B{\bf 10}, 719 (1999). 
     
 
\end{thebibliography}
\end{document}